# A simple rule for axon outgrowth and synaptic competition generates realistic connection lengths and filling fractions

Marcus Kaiser[1,2], Claus C. Hilgetag[3,4], Arjen van Ooyen[5]

[1] School of Computing Science, Newcastle University, Claremont Tower, Newcastle upon Tyne, NE1 7RU, United Kingdom

[2] Institute of Neuroscience, Henry Wellcome Building for Neuroecology, Newcastle University, Framlington Place, Newcastle upon Tyne, NE2 4HH, United Kingdom

[3] Jacobs University Bremen, School of Engineering and Science, Campus Ring 6, 28759 Bremen, Germany

[4] Boston University, Sargent College, Department of Health Sciences, 635 Commonwealth Ave Boston, MA 02215, USA

[5] Department of Integrative Neurophysiology, Center for Neurogenomics and Cognitive Research, VU University Amsterdam, De Boelelaan 1085, 1081 HV Amsterdam, The Netherlands

Corresponding author: Dr. Marcus Kaiser, School of Computing Science, Newcastle University, Claremont Tower, Newcastle upon Tyne, NE1 7RU.
E-mail: M.Kaiser@ncl.ac.uk   Phone: +44 191 222 8161    Fax: +44 191 222 8232

Running title:
Random outgrowth and synaptic competition generates realistic neural connectivity

# Abstract

Neural connectivity at the cellular and mesoscopic level appears very specific, and is presumed to arise from highly specific developmental mechanisms. However, there are general shared features of connectivity in systems as different as the networks formed by individual neurons in *Caenorhabditis elegans* or in rat visual cortex and the mesoscopic circuitry of cortical areas in the mouse, macaque and human brain. In all these systems, connection length distributions have very similar shapes, with an initial large peak and a long flat tail representing the admixture of long-distance connections to mostly short-distance connections. Furthermore, not all potentially possible synapses are formed, and only a fraction of axons (called filling fraction) establish synapses with spatially neighbouring neurons. We explored what aspects of these connectivity patterns can be explained simply by random axonal outgrowth. We found that random axonal growth away from the soma can already reproduce the known distance distribtion of connections. We also observed that experimentally observed filling fractions can be generated by competition for available space at the target neurons—a model markedly different from previous explanations. These findings may serve as a baseline model for the development of connectivity that can be further refined by more specfic mechanisms.

(200 words)

In the wiring length distributions of various neural systems, there is a clear tendency towards low connection lengths, meaning that most neuronal projections are short. Nonetheless, a substantial number of long-distance connections exist, and these connections are important for reducing the number of processing steps in cortical and neuronal networks (Kaiser and Hilgetag, 2006). On several scales of neural systems, from the connectivity between individual neurons in *C. elegans* or the mammalian visual cortex to the connectivity between large-scale cortical regions, the probability of establishing a connection decays exponentially with the distance between neurons or regions.

There are several possible explanations for how this exponential distance dependence may arise. For example, the target of a connection could be genetically encoded by axon guidance molecules (Borisyuk et al., 2008; Yamamoto et al., 2002). Under such a scenario, the target region secretes guidance molecules that diffuse through the tissue. Since there is a higher concentration near the source where the molecule is secreted than at distant regions, a concentration gradient arises, and axons can travel in the direction of higher concentration, towards the target region. The concentration of molecules deposited at one location decays exponentially with time and with distance away from that location (Murray, 1990). This relationship is given by $c(x,t) = \frac{Q}{2\sqrt{\pi D t}} e^{\frac{-x^2}{4 D t}}$ where $D$ is the diffusion coefficient and $Q$ is the initial amount of particles per area. Given a threshold for the detection of guidance molecules, neurons closer to the source will be more likely to pick up the signal than neurons further away, leading to distance dependence in the establishment of connections. However, this model would incur a hard border – a distance beyond which the concentration falls below the threshold and where no connections occur, whereas for distances shorter than the critical distance, the probability of establishing a connection would be close to 100%. Under this regime, an inverse sigmoid distribution of connection lengths would be expected, which is unlike the curves that are found in neural systems. In addition, guidance through attraction and repulsion is usually discussed for the global level of long-distance connectivity but not for the local connectivity within regions. Whereas guidance cues are a defining factor at the global level, their role on the local level is limited, due to the problem of establishing a stable gradient for very short distances (less than 0.7 mm). The standard hypothesis for guidance clues is that they are provided by controlled gene expression, specific transmitter-receptor systems and growth factors (Sperry, 1963; Yamamoto *et al.*, 2002). This hypothesis involves diffusion-based mechanisms. Recent studies investigated the genetic component of connection development in *C. elegans* (Baruch et al., 2008; Kaufman et al., 2006), and found that a substantial amount (ca. 40% on average) of the variability of connection patterns in *C. elegans* cannot be accounted for by differences in gene expression patterns (Baruch *et al.*, 2008; Kaufman *et al.*, 2006).

An alternative hypothesis, called 'Peters Principle' by Braitenberg and Schuez (Braitenberg and Schuez, 1998), suggests that neural outgrowth is basically random, and that specificity in the wiring is derived from the overlap of specific neuronal populations (Binzegger et al., 2004). Here we explore whether basic mechanisms of random axonal growth can also explain the connection length distributions in neural networks. We focus primarily on local connectivity and assume a simple model with unspecific establishment of connections.

It is well known that axonal growth cones (specialized structures at the end of outgrowing neurites) tend to grow in approximately straight lines unless their way is obstructed or when they encounter attractive or repellent guidance cues in the surrounding medium (Easter et al., 1985; Sperry, 1963; Yamamoto *et al.*, 2002). Given a uniform distribution of neurons in a two dimensional space, a straight growing axon is more likely to hit a neuron in its immediate vicinity than further away. If we assume that the axon establishes a connection with the first neuron that it encounters, the probability of establishing connections between neurons would decay exponentially with distance. A further condition that we test is that neurons with which the axon could make a connection are already saturated with other incoming connections. That is, the dendritic tree of the neuron is already occupied with input synapses from other neurons. In that case, the axon would need to continue its growth along the same direction in order to find another candidate target. This aspect of the model introduces the idea that the number of synapses that can be established on a dendrite is limited. This occurs if the dendritic surface area and consequently the number of spines is limited.

Although numerous models for the establishment of neural connections have been formulated (Van Ooyen A., 2005; Van Ooyen A. et al., 1995; van Ooyen A. and Willshaw, 1999), and models exist that describe the exponentially decaying connection probability (Amirikian, 2005), to our knowledge no model has been presented for the underlying mechanisms by which this distance dependence in connectivity can arise.

In addition to providing a model for distance-dependent connectivity, here we also show how occupation of neurons introduces competition for connection targets. As a result of such competition, only a fraction of all possible synapses (actual connections) is formed, out of the many potential connections that might occur when an axonal growth cone is spatially close to a dendritic tree (Figure 1). This so-called filling fraction (Stepanyants et al., 2002) has been measured experimentally and is found to vary with the maximum number of synapses per connection or per neuron. In brief, our model reproduces the widely observed connection length distributions through random neural outgrowth, and provides an explanation of experimentally found filling fractions through the additional factor of neural target competition. Naturally, there are several other topological and spatial network properties (Costa Luciano da Fontoura , Kaiser et al., 2007) that can be considered, such as characteristic path lengh, clustering coefficient, or modular organisation (Costa L. d. F., Rodrigues et al., 2007). In the present study, however, we focus on filling fraction and length distribution, because these comprise network features that are most often reported in anatomical studies. However, when more detailed neuronal connectivity data become available, one can readily compare model and experimental data using a variety of additional connectivity measures.

## Materials and Methods

### *Neural networks*
**Primate corticocortical network**

We analyzed the spatial arrangement of 2,402 projections among 95 cortical areas and sub-areas of the primate (Macaque monkey) brain. The connectivity data were retrieved from CoCoMac (Kötter, 2004; Stephan K. E. et al.) and are based on three extensive neuroanatomical compilations (Carmichael and Price, 1994; Felleman and van Essen, 1991; Lewis and Essen, 2000) that collectively cover large parts of the cerebral cortex. Spatial positions of cortical areas were estimated from surface parcelling using the CARET software (http://brainmap.wustl.edu/caret). The spatial positions of areas were calculated as the average surface coordinate (or centre of mass) of the three-dimensional extension of an area (cf. Kaiser and Hilgetag, 2004).

*C. elegans* **neuronal networks**

We further analyzed two-dimensional spatial representations of the global neuronal network (277 neurons, 2,105 connections) of the nematode *C. elegans*, as well as a local sub-network of neurons within *C. elegans* rostral ganglia (anterior, dorsal, lateral, ring, 131 neurons; 764 unidirectional connections). Spatial two-dimensional positions (in the lateral plane), representing the position of the soma of individual neurons in *C. elegans*, were provided by Y. Choe (Choe et al., 2004). Neuronal connectivity was obtained from (Achacoso and Yamamoto, 1992). This compilation is largely based on the data set of White et al. (White et al., 1986) in which connections were identified by electron microscope reconstructions. Details of the data set are described in Kaiser and Hilgetag (2006). Wiring length was calculated as direct Euclidean distance between connected components in 3D (Macaque) or 2D (*C. elegans*). Both datasets are available at http://www.biological-networks.org.

**Rat local network**

The data was taken from Lohmann & Roerig (Lohmann and Rorig, 1994). It contains connections of supragranular pyramidal neurons (layers II and III) in the rat extrastriate visual cortex. The study measures the length of 467 horizontal axon collaterals in area 18a. Only 3-5 of the longest branches of neurons were examined by light microscopy (Lohmann and Rorig, 1994).

## *Neural connectivity simulation*
**Sparse neuron population**

We tested a condition in which 400 neurons were randomly arranged on a 100x100 unit two-dimensional embedding space. As each cell filled one square unit, only 4% of the potential space was filled with neuron bodies. Subsequently, each neuron was randomly assigned one direction for axon outgrowth. The axon was growing in a straight line until another neuron was hit; thereafter, a connection between both neurons was established. The straight axonal growth away from the soma reflects the experimentally observed somatofugal growth pattern of neurites (Samsonovich and Ascoli, 2003). The axon continued growing until the borders of the embedding space were reached or the maximum number of established synapses for that neuron had been reached. An axon was considered to be close enough to another neuron for establishing a connection when the Manhattan distance between the axon tip or growth cone and the neuron was smaller than one unit. The projection length of the established fibre was given as the Euclidean distance between the two connected neurons. The morphology of dendrites was not included in the distance calculation as the model should be as simple and general as possible for all types of neural tissue. Similarly, axon collaterals were not taken into account to ensure that our results were comparable to other modelling studies and to reduce the number of assumptions of the model.

Under what we call the occupied condition, a neuron was considered unable to establish a connection with a target neuron when the target already had another incoming connection. In such cases, the neuron could not accommodate any additional synapses. As in the previous condition, axons were growing in a random direction until reaching a potential target neuron. If that neuron was occupied—that is, when it already had an incoming connection from another neuron— the axon continued to grow along the same direction until a connection with an unoccupied neuron could be established (or the axon left the embedding space for network growth). We used an upper bound of one incoming connection per neuron as a default but also tested other values (see Figure 1 for a schematic overview of competition of incoming connections).

**Dense neuron population**

In the so-called dense condition, 1000 neurons were placed on the 100x100 unit field, filling 10% of the space. Again, the axon was growing in a straight line in a randomly chosen direction and a connection was established with the first neuron it encountered along the way. Also in this scenario, we tested both the potentially occupied and unoccupied condition.

**Evolution of neuron population size**

Under this condition, the simulation started with one neuron positioned randomly within the 100x100 unit embedding field and neurons were subsequently added at each time step until the resulting network reached a density of 1000 neurons. The connections of a neuron were established once it was added to the network. Note that connections could only be established to neurons that existed at that time step. Again, we tested both the potentially occupied and unoccupied condition.

**Filling fraction**

The filling fraction of a neuron is the number of actually established synapses divided by the number of potential synapses (Stepanyants *et al.*, 2002). Potential synapses can be formed by all axons that are within reach of a neuron. In our simulation this meant that they were less than one spatial unit away from the centre of the neuron. In the following, the filling fraction represents the average value over all neurons as measured at the end of the simulation.

# Results

## *Wiring Length Distributions of Neural Networks*

In the primate corticocortical as well as in the rat and *C. elegans* neuronal network, the reach of connections among cortical areas or neurons quickly decays with distance (Figure 2 A-D). Nonetheless, some connections are not formed between immediate neighbouring cortical areas and extend over a considerable distance (primate cortical network, Figure 2 A). This is also true for the local connectivity among neurons within cortical areas (Figure 2 B). A similar distribution emerged for the local connectivity in *C. elegans* (Figure 2 C). For the global *C. elegans* network, some connections are almost as long as the entire organism (Figure 2 D). Thus, all neural systems show a tendency toward short-distance projections while still allowing for the existence of long-distance projections.

We fitted the distance distributions with exponential ($a \cdot e^{bx}$), power-law ($a \cdot x^b$), Weibull ($a \cdot b \cdot x^{b-1} \cdot e^{-ax^b}$), Gaussian ($a \cdot e^{-\left(\frac{x-b}{c}\right)^2}$), and Gamma ($c \cdot f(x,a,b)$) distributions. Note that the Gamma distribution used an additional factor *c* for scaling the probability density function *f*. The functions as well as the $R^2$ as a measure of the goodness of the fit are shown in Table 1. Because the observed anatomical distributions were non-symmetrical, we chose the Gamma distribution for evaluating the anatomical length distributions as well as for the distributions resulting from simulated neural development. The fitting parameters of the anatomical networks for the Gamma distribution are given in Table 2.

## *Wiring Length Distributions of Generated Networks*

**Sparse Neuron Population**

On a 100x100 unit grid, 400 neurons were randomly positioned. Axons that hit a target, i.e. that had a Manhattan distance of less than one unit at that growth step, were establishing a connection on the first encounter, irrespective of there already being other incoming connections at the target (Figure 3 A, B). We also tested the condition where targets could not receive connections once they already had an incoming axon (Figure 3 C, D). In this case, we found that the number of established connections was reduced and connection lengths were longer. In the first condition, connections were on average about 16.4 units long (maximum length: 95 units), whereas in the second condition connections were on average 22.6 units (maximum length: 114 units) long.

**Dense Neuron Population**

On a 100x100 grid, 1000 neurons were randomly positioned. Axons that hit a target established a connection on the first encounter, again irrespective of there already being other incoming connections (Figure 4 A, B). We also tested the condition where targets could be occupied when they already had an incoming axon (Figure 4 C, D). In this case, the number of established axons was reduced and average connection lengths were much longer. In the first condition, connections were on average about 10 units long (maximum length: 102 units), whereas in the second condition, connections were on average as long as 18 units (maximum length: 101 units). Note that in a dense neuronal population the effect of potential occupation on the average connection length was considerably stronger than in the sparse population.

In the spatially dense neuron population, the number of connections scaled with the number of neurons (that is, with a factor of 2.5 for an increase from 400 to 1,000 neurons). However, the average connection length was lower than for the sparse network. This was due to the fact that in the vicinity of each node there were more neurons with which a connection could be established.

**Evolution of neuron population size**

We also tested the case in which the 1,000 neurons were added one after another, starting with one initial neuron (Figure 5 A, B). Again, we looked at the effect of occupation where neurons that had already received one axon were unavailable for further incoming connections (Figure 5 C, D). The average connection lengths were comparable to the non-evolving network with 1,000 nodes with average connection lengths of 9.7 units (no competition) and 18.4 units (competition).

For the case with potential occupation, fewer connections were established. This is due to competition for free places, whereby some axons did not successfully establish a connection before reaching the borders of the embedding space. Because some neurons in the axonal pathway would already be occupied, the length of successfully established connections increased. There seemed to be no difference in this respect with the other two cases with potential occupation, indicating that competition for free places and neuron density rather than a static versus evolving neuron population influenced connection length distributions. In the following sections, the role of competition on synapse formation is investigated more closely.

### *Role of competition for available neuronal targets*

The filling fraction is the number of established synapses divided by the number of potential synapses. A potential synapse occurs whenever axon and dendrite are spatially close (Stepanyants *et al.*, 2002). The filling fraction took a value of one if all potential synapses were established, which occurred in our simulations without potential occupation. In the simulations that included potential occupation, the filling fraction decreased with increasing numbers of neurons (Figure 6 A). Previous work pointed out that a low filling fraction increases the number of ways in which synapses could be (re)distributed among target spaces, leading to maximum plasticity whereas wiring should be as short as possible (Chklovskii, 2004). Simulations also showed a decrease in the probability that any two neurons were connected (edge density; Figure 6 B) when the total number of neurons increased. For a low number of neurons, the edge density of the whole network differs from the edge density when isolated nodes are not taken into account. This shows that isolated nodes, i.e., nodes that neither establish nor receive a synaptic connection, occur frequently when the number of potential targets in the axonal pathway is low, but are rare at later stages of development when the number of neurons is higher.

Such a relative decrease in the number of connections occurs both in the whole network and when only non-isolated nodes with at least one edge are taken into account. This indicates that with an increasing number of competing neurons, many axons reach the borders of the embedding space without establishing a connection with any other neuron.

For the previous simulations with 1,000 neurons, there was no difference in filling fraction, neither between constant or variable cell sizes (see section *Robustness of results*) nor between a static or evolving number of neurons (the overall filling fraction remained around 0.26). Note that an increase in the number of neurons has a linear relation with the spatial, or neuron, density. Therefore, the filling fraction decreases with neuron density. This relation was also found in experimental studies on connectivity in mouse, the rat hippocampus, and the macaque visual cortex where the filling fraction ranges from 0.12 to 0.34 (Stepanyants *et al.*, 2002).

If several (here: 10) potential places for synapse formation exist on the neuron, axonal growth cones form synapses earlier so that fewer long-distance connections are established (Figure 7). Since ample opportunities for synapse formation exist, all potential synapses can be established and the filling fraction becomes 1, whereas it is 0.28 with only one available place.

A simultaneous increase of synapses and available space leads to a gradual increase in filling fraction with the number of synapses or places, from 0.28 for one synapse to 0.47 for eight synapses (diagonal of Figure 8). Each neuron is growing and establishing synapses until the maximum number of outgoing synapses for that neuron or the borders of the embedding space have been reached. We also tested scenarios where the number of outgoing synapses and of available targets differed (Figure 8): Having more outgoing synapses than free places led to more competition and therefore to a lower filling fraction (<0.2). More free places than outgoing synapses, on the other hand, reduced competition and led to a higher filling fraction (>0.6). Testing different scenarios showed that the filling fraction *f* depended on the competition factor *r* (ratio between outgoing synapses and maximum free places) following an exponential decay, $f(r) = 1.125 \exp(-r)$, as shown in Figure 9.

**Robustness of results**

*Variable cell size:* In the previous simulations, the size of cells was homogenous with a diameter of one unit, but we also tested variability in cell size. For the variable-cell-size condition, the size was randomly chosen from a uniform distribution [0; 2] resulting in an average size of one unit. We found that these individual differences in cell size did no lead to any significant changes in filling fraction or connection length distribution (see Supplementary Material; Figure S1).

*Growth in three dimensions:* In the results presented so far we have focused on two-dimensional growth which reflects the situation for *in vitro* slice experiments or *in vivo* growth where one dimension is physically absent. However, we also tested growth in three dimensions which led to a similar shape of connection length distributions and similar filling fractions (see Supplementary Material; Figures S2 and S3).

# Discussion

We have tested three scenarios of synapse formation during cortical development: (1) axon growth in a spatially sparse versus a spatially dense neuronal population, (2) a gradual increase in the number of neurons versus all neurons present from the outset of development, and (3) having a limit on how many incoming connections a neuron can receive versus a potentially unlimited number of incoming connections.

In all modelled scenarios, the frequency of connections decayed with the distance between neurons. This decay with distance could be fitted with a Gamma distribution, with an excellent goodness of fit (adjusted $R^2 \geq 0.98$). In analogy, experimental data for the anatomical networks have shown a closely related Gaussian distribution (Hellwig Bernhard, 2000). Furthermore, whereas the gradual increase in the number of neurons (scenario 2) led to fewer connections in the network, the connection lengths increased because some neurons were unavailable for accommodating incoming connections.

We also introduced the idea that the filling fraction—the proportion of established synapses relative to all possible synapses—may arise from neural competition for axonal space on the dendritic tree, rather than being the result of a hit-and-miss model of axon growth. Our model can reproduce filling fractions that were experimentally found in anatomical networks and leads to testable predictions for the development of network connectivity. For example, a decrease in axon collaterals or an increase in space on target neurons should reduce competition and therefore lead to higher filling fractions. Conversely, smaller dendritic trees, with the neuron density remaining the same, should lead to reduced filling fractions. Indeed, a relationship has been found between the size of the dendritic tree and the number of innervating axons that survive into adulthood (Purves and Lichtman, 1980), pointing to competition for space in the developmental pruning of axonal connections.

**Data choice**

We used anatomical data from the global level of cortical fibre tracts down to the local level of neuronal connectivity in the rat visual cortex and *C. elegans*. These datasets are the best available but not the best imaginable datasets: The primate corticocortical network reflects macroscopic connectivity, whereas our model concerns microscopic network growth. *C. elegans* is a system affected by both small size and a ganglionar organisation, which is different from the three or six-layer laminar organisation of other neural systems. The rat data set, the best proxy of local neuronal networks, is incomplete as only some layers and only the longest branches of connections are reported. However, other more relevant datasets were unavailable to us and the datasets used in this study represent the best datasets we could find. Such limitations of incomplete datasets are typical of local anatomical studies and it would be very useful for future studies if different datasets at the local level could be combined in similar ways as cortical fibre tract data have been combined (Kötter, 2004; Stephan K. E. *et al.*, 2001; Stephan K E et al., 2000).

**Network topology**

Although the connection length distributions in the generated networks were similar to those found in anatomical networks, modelled and biological networks differed in their topological organisation. We tested whether the generated networks showed features of small-world networks (Watts and Strogatz, 1998), as found previously in neural connectivity as diverse as the neuronal network of *C. elegans* and the cortical networks of the cat and macaque brain (Hilgetag et al., 2000). These networks show a low characteristic path length that is comparable to that in a randomly organized network. The characteristic path length denotes the average shortest number of connections that have to be crossed in order to go from one neuron or cortical area to another. At the same time, the clustering coefficient, denoting how well direct neighbours of a node are on average connected, is much higher than for random networks (for a genal overview of network features of neural systems, see Sporns et al., 2004). In our simulations, the characteristic path length of the generated networks was below the one for random networks in five out of the six scenarios (it was higher for the non-evolving dense matrix without potential occupation), which means that nodes of the networks can be reached in fewer steps than in a comparable randomly organized network. The clustering coefficient, however, was above the value for random networks in all cases. Additional mechanisms must therefore account for the observed small-world and multiple-cluster organisation of biological neural systems. One such mechanism is the occurrence of time sequences and time windows during development: Some parts of the network will develop earlier than others and the connections will preferentially be established with neurons that are in the same time-window for synaptogenesis (Kaiser and Hilgetag, 2007; Nisbach and Kaiser, 2007). Examples would be the time course for the establishment of layers within cortical areas or time-courses of synaptogenesis and neuron migration (Rakic Pakso, 2002).

We could also have considered additional spatial and topological network properties (Costa Luciano da Fontoura , Kaiser *et al.*, 2007), but possibilities for comparison with experimental data are limited because of the limited knowledge of network connectivity at the local neuronal level. Such measures include the degree distribution (Kaiser et al., 2007), the spatial layout of neurons (Kaiser and Hilgetag, 2006), and various other network properties (Costa L. d. F., Rodrigues *et al.*, 2007). Future studies might thus show that our model agrees with some properties of neuronal networks but not with others.

**Model limitations**

The present study uses a simple model that ignores many details of neural development, in order to establish what aspects of connectivity may be explained by a minimal mechanism of random axonal growth. If general features of connectivity, such as the typically observed wiring length distribution, can already be explained by random growth, one may conclude that they do not require intricate developmental mechanisms—for instance, axon guidance signalling between pre- and postsynaptic neurons. Such a finding then shifts the focus of research to connectivity features for which more specific mechanisms are indeed required.

Clearly, several aspects of connectivity formation are not random, as demonstrated by the orderly collection of axonal fibers into nerve fiber tracts. The intricate vertical and laminar organization of cortical connections may also be shaped by specific factors, including patterns of gene expression (Tarui et al., 2005), that might affect connectivity formation. Moreover, the geometrical shapes of dendritic and axonal arborizations, with their characteristic curvature, branch angles, tree asymmetry, fractal dimensions, etc. (for a review, see Uylings and van Pelt, 2002) may also be important for determining connectivity patterns and space filling (Wen and Chklovskii, 2008). It is also likely that there are differences between species (Easter *et al.*, 1985) as well as between spatial scales—for example, the formation of connections between individual neurons or between large-scale cortical areas. While the assembly of mesoscopic connections may proceed largely in the absence of neural activity (Verhage et al., 2000), synaptic connectivity of intrinsic connections is adjusted via activity-dependent plasticity (Butz et al., 2009). Nonetheless, as our model demonstrates, basic features of connectivity could be formed by random growth processes at the cellular as well as systems level.

**Random outgrowth and competition for space**

In our model, we assumed a random outgrowth of axons. This assumption is in accordance with experimental results showing that synapse formation, within a specified class of target neurons, can be described as a random process. For example, the distribution of synaptic boutons on a dendrite as well as on pyramidal cell collaterals can be modelled by a Poisson process (Hellwig B. et al., 1994). In addition, the connectivity pattern of layer-III pyramidal neurons in cat and monkey visual cortex agrees with the hypothesis of random wiring (Kisvarday et al., 1986; McGuire et al., 1991).

We also assumed that neurons might become unavailable for receiving further connections once they already possess a sufficient number of incoming connections. This mechanism could provide an explanation for the experimental observation that axons and dendrites may sometimes be spatially adjacent without establishing synaptic contacts. For example, confocal microscopy and recordings in L5 pyramidal neurons of rat somatosensory cortex have shown that less than one quarter of touches between connected neurons led to the formation of a synapse. In total, only 10% of all possible synapses (filling fraction, see Stepanyants *et al.*, 2002) for adjacent axons and dendrites were established (Kalisman et al., 2005). In cases where a connection cannot be established, axons will continue to grow beyond the spatially nearby neuron to establish connections with neurons further away.

During the development of the nervous system, cells are commonly innervated by more axons than they ultimately maintain into adulthood (LaMantia and Rakic, 1990; Lohof et al., 1996; Purves and Lichtman, 1980; Rakic P. et al., 1986; van Ooyen A., 2001). This is a widespread phenomenon in the developing nervous system and occurs, for example, in the formation of ocular dominance layers and columns in the LGN and visual cortex (e.g., Wiesel, 1982), the climbing fibre innervation of Purkinje cells (Crepel, 1982), the innervation of sympathetic and parasympathetic ganglion cells (Purves and Lichtman, 1980), and the development of connections between motor neurons and muscle fibres (e.g., Sanes and Lichtman, 1999). Importantly for our model, the overproduced axons become eliminated by competitive interactions between the innervating fibres (Rakic P., 1981; Rakic Pasko, 1986; Rakic P. and Riley, 1983), and may involve competition for target-derived neurotrophic factors (Cabelli et al., 1995). Interestingly, also the cellular geometry appears to play an important role. In many types of neurons, a positive correlation exists between the size of the dendritic tree and the number of innervating axons surviving into adulthood (Hume and Purves, 1981; see also van Ooyen A. et al., 2000).

**Other models for filling fraction**

A model for the filling fraction was first described by Stepanyants et al. (2002). In this model, non-establishment of synapses was explained by axons running close to the dendritic tree of a neuron but missing the dendritic spine in their straight-forward movement. However, whereas this model numerically reproduced the experimentally found filling fraction, the explanation may not be biologically realistic: Filopodia of axonal growth cones can be up to 40 μm in length (van Ooyen Arjen, 2003) potentially covering up to 80 μm in the forward direction searching for potential binding places. The distance between spines on the dendritic tree, however, is significantly smaller than this search space, with typically two spines/μm for dendritic segments of cortical pyramidal cells (Braitenberg and Schuez, 1998). Competition might provide a more realistic explanation for the experimentally observed filling fractions.

**Comparison with local cortical connectivity**

If some aspects of neuronal connectivity at the local level can result from non-specific axonal outgrowth, as suggested by the present model, we would predict for neurons that develop early during development that they (1) establish relatively more short-distance connections compared to later developing neurons and (2) tend to establish more connections. Note, however, that the second effect might be limited, because, although few neurons are already occupied early in development, the number of neurons, and therefore potential partners, is lower as well. Our results concerning the relation between the filling fraction and the competition factor for establishing synapses could be tested experimentally for local neuronal connections.

**Conclusion**

Simple models that assume a straight outgrowth of axons in a randomly chosen direction can account for the exponential decay in the connection length distribution that is found experimentally in diverse neural networks in the primate and rat brain as well as the nematode *C. elegans*. In addition, spatial occupation of postsynaptic neurons and axonal competition for available synaptic targets leads to filling fractions that are similar to those found experimentally. The model is a simple starting point for modelling neuronal network development and may turn out to be incomplete when compared with anatomical datasets that can be characterised more extensively by a larger set of network parameters. In such a case, some of the factors that are now not part of our simple model would need to be included. We hope that this study will encourage both the modelling of neuronal network development as well as anatomical measurements of neuronal connectivity.

**Funding**

This work was supported by the Engineering and Physical Sciences Research Council [EP/E002331/1 and EP/G03950X/1 to M.K.] and the Royal Society [RG/2006/2 to M.K.].

**Acknowledgements**

We thank Yoonsuck Choe for providing us with spatial position data of *C. elegans* neurons and Miles Whittington and Evelyne Sernagor for helpful comments on the manuscript.

# Tables

**Table 1.** Performance of different fitting functions for anatomical length distributions. The table shows the different distributions and $R^2$ as a measure of the goodness of fit (see Table 2 for the definition of Gamma used here).

| Network | $R^2$ | | | | |
| --- | --- | --- | --- | --- | --- |
| | Exponential | Power-law | Weibull | Gaussian | Gamma |
| **Macaque cortex, cortical** | 0.609 | 0.348 | -0.739 | 0.879 | 0.960 |
| **Rat visual cortex, neuronal** | 0.080 | 0.001 | 0.024 | 0.927 | 0.900 |
| ***C. elegans*, local neuronal** | 0.598 | 0.303 | 0.305 | 0.910 | 0.957 |
| ***C. elegans*, global neuronal** | 0.974 | 0.981 | 0.979 | 0.974 | 0.824 |

**Table 2.** Gamma fit for different anatomical networks (*f* denotes the Gamma probability density function with shape parameter *a* and scaling parameter *b*). Parameters *a, b* and *c* relate to the relative length frequency *P(l)* where *l* is the fibre length in mm.

| Gamma fit: $P(l) = c \cdot f(l,a,b)$ | a | b | c |
| --- | --- | --- | --- |
| **Macaque cortex, cortical network** | 3.078 | 6.378 | 5.783 |
| **Rat visual cortex, neuronal network** | 6.075 | 0.079 | 0.151 |
| ***C. elegans*, local neuronal network** | 2.486 | 0.010 | 0.009 |
| ***C. elegans*, global neuronal network** | 0.541 | 0.419 | 0.210 |

# Captions

**Figure 1. Competition for establishing neural connections**. (*A*) An axonal growth cone can establish a synaptic connection with a dendritic spine if there is available space. (*B*) If all available dendritic spaces are already occupied by competing neurons, the potential synapse will not be established and the axon will continue its search and eventually establish connections with other neurons in its path.

**Figure 2. Connection lengths in cortical and neuronal networks.** Connection length distributions (x) fitted with a Gamma function (see Table 2 for fit coefficients). (*A*) Macaque cortical network. (*B*) Rat supragranular neuronal network. (*C*) *C. elegans* local network. (*D*) *C. elegans* global network.

**Figure 3. Spatially sparse population.** (*A*) Network of 400 nodes and 264 unidirectional connections. (*B*) Distribution of connection lengths (x) with the Gamma fit indicated as a line (a = 1.5; b = 11.1; c = 11.3; $R^2$ = 0.96). (*C*) Network of 400 nodes with potential occupation and 217 unidirectional connections. (*D*) Distribution of connection lengths (x) with the Gamma fit indicated as a line (a = 1.5; b = 15.0; c = 12.6; $R^2$ = 0.98).

**Figure 4. Spatially dense population.** (*A*) Network of 1000 nodes and 796 connections. (*B*) Distribution of connection lengths (x) with the Gamma fit indicated as a line (a = 1.5; b = 6.8; c = 10.5; $R^2$ = 0.97). (*C*) Network of 1000 nodes with potential blocking and 690 unidirectional connections. (*D*) Distribution of connection lengths (x) with the Gamma fit indicated as a line (a = 1.5; b = 12.5; c = 11.7; $R^2$ = 0.97).

**Figure 5. Evolution of population size**. (*A*) Network of 1000 nodes and 809 connections. (*B*) Distribution of connection lengths (x) with the Gamma function fit indicated as a line (a = 1.5; b = 6.4; c = 10.0; $R^2$ = 0.97). (*C*) Network of 1000 nodes with potential occupation and 697 unidirectional connections. (*D*) Distribution of connection lengths (x) with the Gamma fit indicated as a line (a = 1.5; b = 12.6; c = 12.1; $R^2$ = 0.98).

**Figure 6. Dependence of filling fraction and edge density on the number of neurons**. (*A*) Filling fraction, which is the proportion of successfully established connections after axon-cell contact, depends on the number of neurons (vertical bars indicate the standard deviation while lines show the average over 10 generated networks). (*B*) The edge density–the number of existing edges in the network (i.e., connected neuron pairs) divided by the number of all possible edges–decreases for larger numbers of neurons (average and standard deviation as in (*A*); solid line: complete network; dashed line: network without isolated nodes).

**Figure 7. Histogram of connection lengths**. Connection lengths were measured for 10 generated networks resulting in the average distributions shown here. (*A*) Maximally one afferent synapse on a neuron. (*B*) Maximally ten afferent synapses on a neuron. The longest possible connection in the embedding space would be 140 units long. Note that due to increased competition for free places, there were more long-distance connections up to 100 units in scenario (*A*) whereas connections in (*B*) where only up to 20 units long.

**Figure 8. Variation of axon collaterals (outgoing synapses) and available places for synapse establishment**. The median filling fraction of 20 generated networks with 1,000 neurons for each parameter pair is indicated by the grey level. The maximum number of outgoing synapses denotes how many connections a single neuron can maximally establish. The number of maximum free places for synapses shows how many places each neuron has available for incoming connections.

**Figure 9. Influence of competition on the filling fraction**. The x-axis shows the ratio between outgoing axon collaterals and maximally available places at potential target neurons of Figure 8. The filling fraction decays with this ratio and follows an exponential function $f(r) = 1.125 \exp(-r)$, $R^2$=0.91 (solid line).

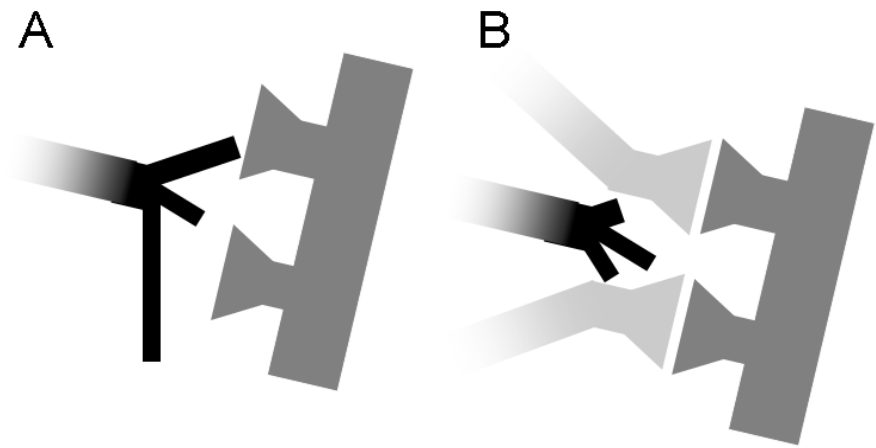

Figure 1

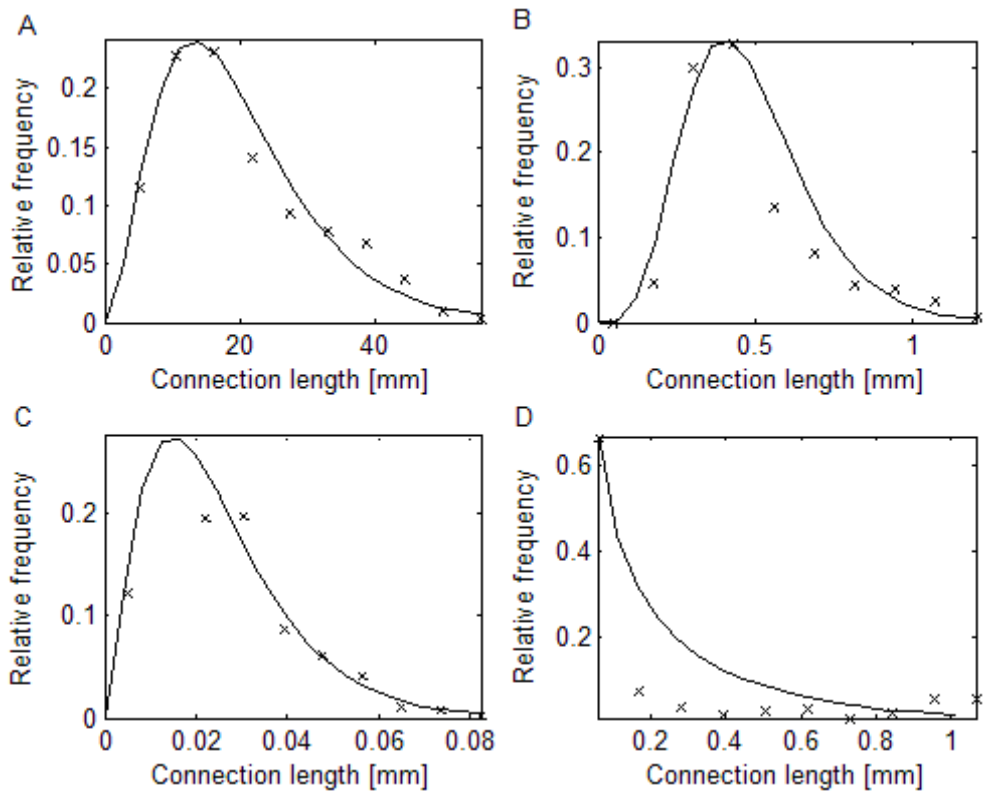

Figure 2

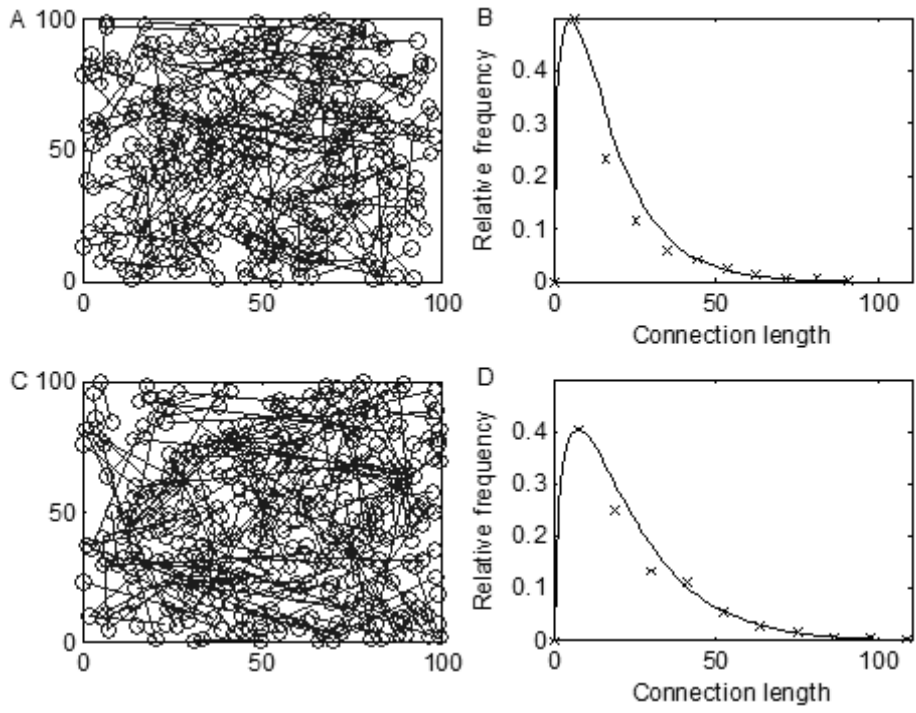

Figure 3

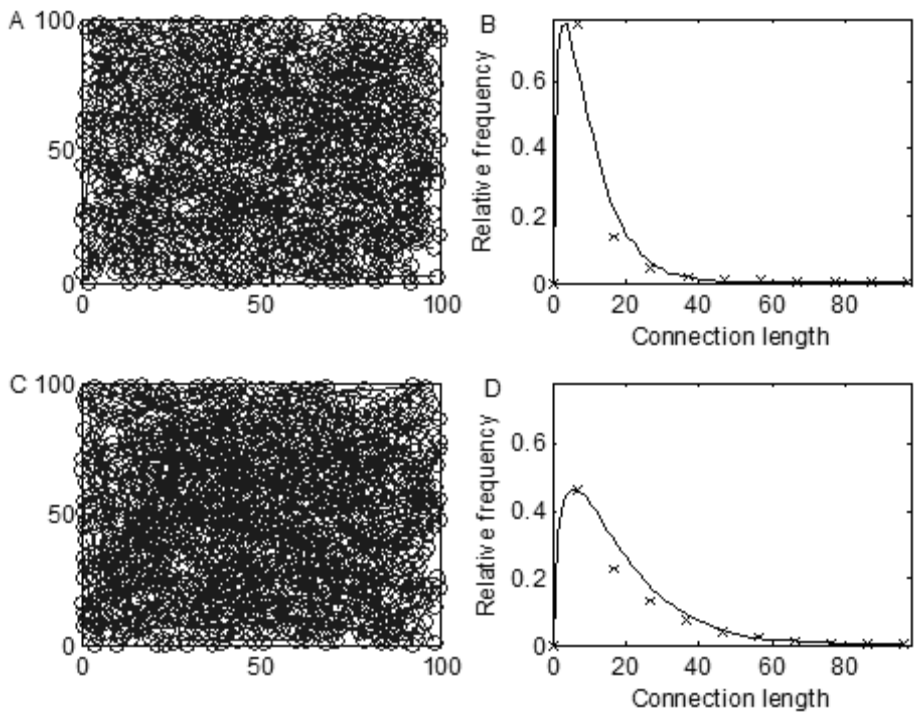

Figure 4

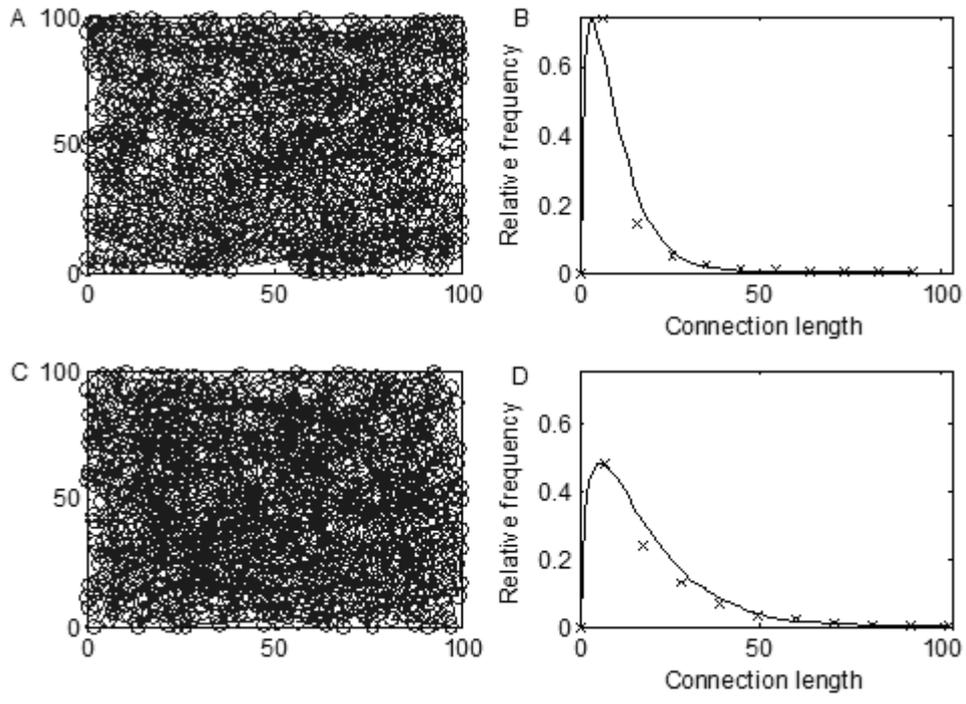

Figure 5

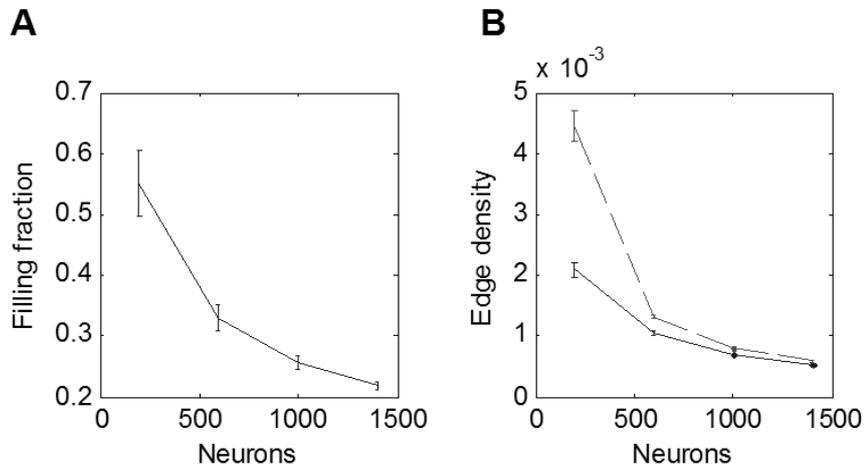

Figure 6

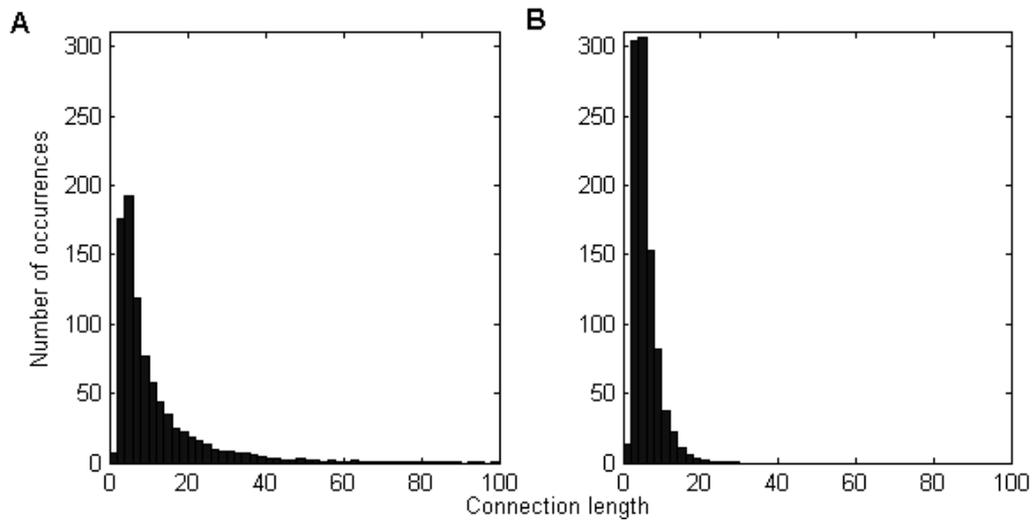

Figure 7

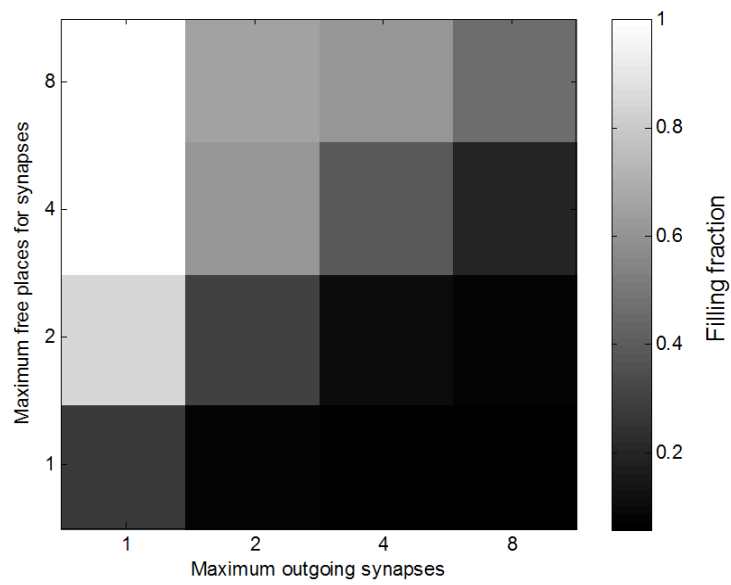

Figure 8

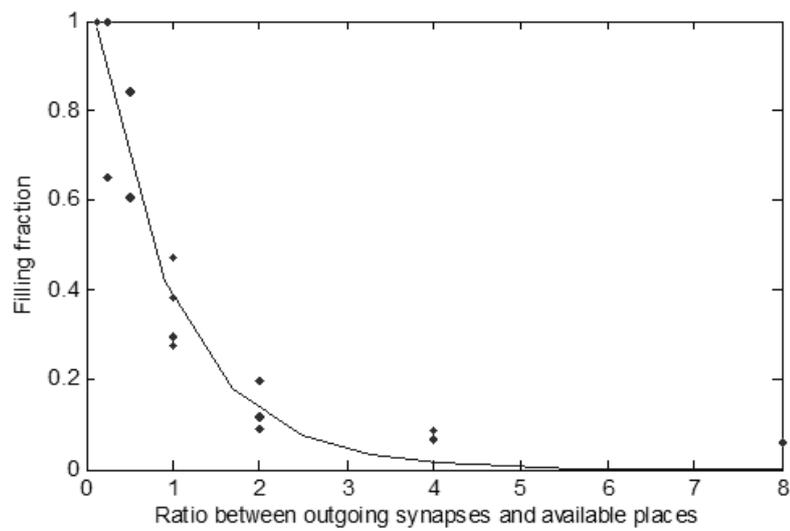

Figure 9

# A simple rule for axon outgrowth and synaptic competition generates realistic connection lengths and filling fractions

Marcus Kaiser, Claus C. Hilgetag, Arjen van Ooyen

## Supplementary Material

**Connection length distributions**

Distribution of connections lengths for different parameter sets for two-dimensional growth of 1000 or 400 neurons. The subscripted label *abc* (e.g., w2d1000$_{abc}$) indicates the growth model with 100 nodes, $a=1$ for static growth, $b=1$ for the number of connections per neuron being limited to one, and $c=1$ if the cell size is variable between 0 and 2 spatial units (compared to the uniform cell size of one in the standard simulations).

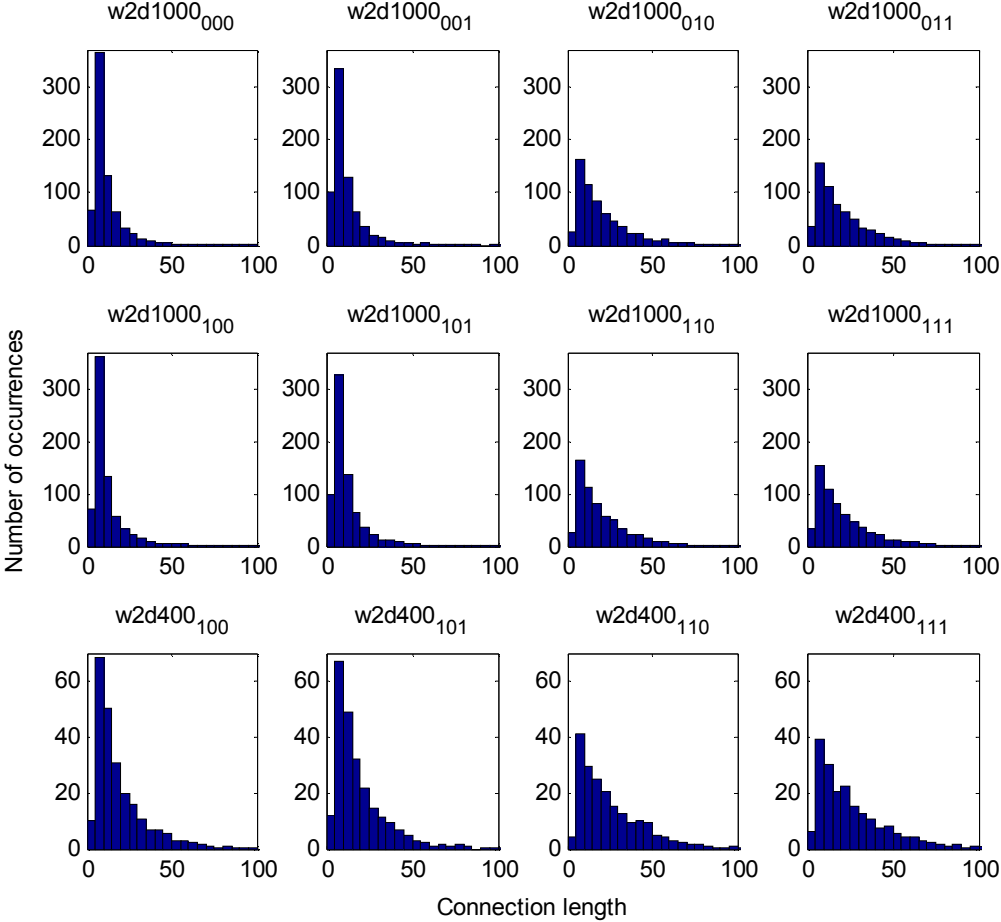

**Figure S1.** Connection length distributions (average over 10 generated networks).

**Filling fraction and connection lengths for three-dimensional axon growth**

We also simulated axonal growth in three dimensions. The embedding space was 34x34x34 units large so that neurons could easily be placed without an overlap of their three-dimensional shape. Figure S2 shows the connections length distributions for 400 and 1000 generated neurons following the same parameters as in Figure S1.

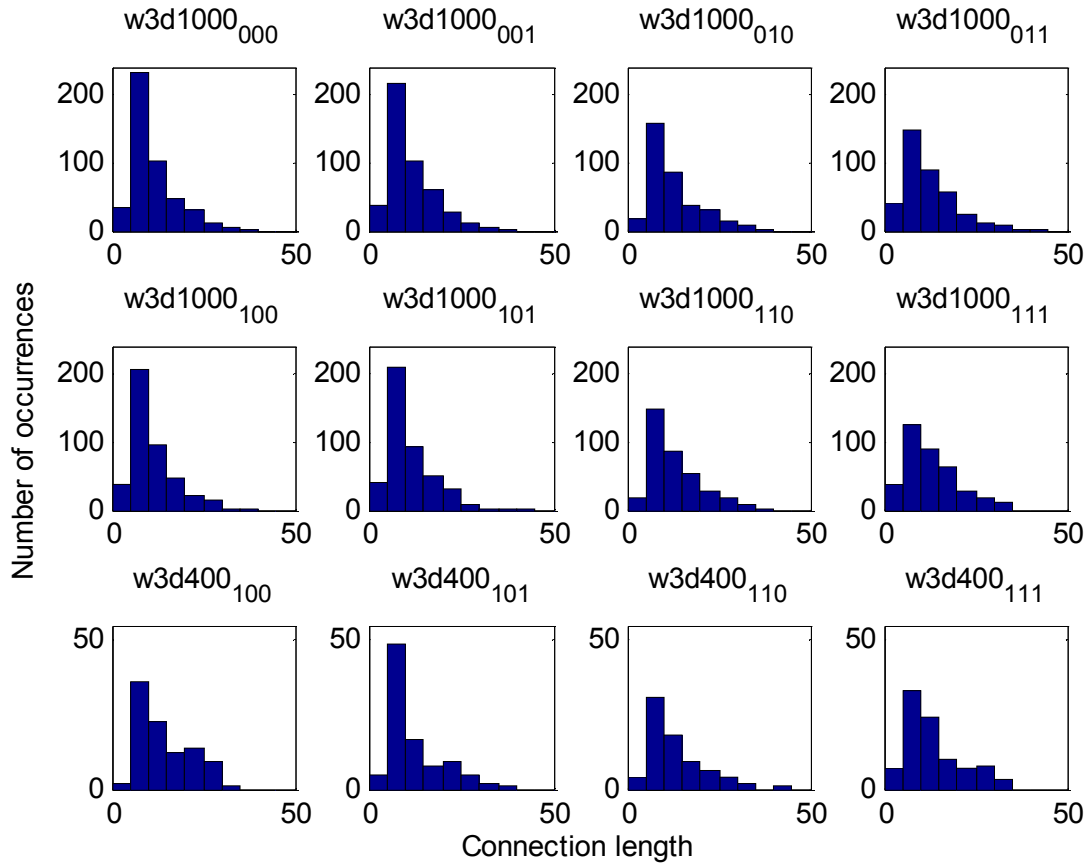

**Figure S2. Connection length distribution for three-dimensional growth.** Average over 10 generated networks; for the description of labels, see Figure S1.

We again tested the relation between the number of neurons (neural density) and the filling fraction and edge density of the resulting three-dimensional networks as we did for figure 6 of the main text (Figure S3). The filling fraction decreases with the neural density (Fig. S3A) as it did for the two-dimensional embedding. Note that the higher absolute filling fractions, compared to two-dimensional growth, are due to lower spatial density (for 1000 neurons, for example, 2.5% of the three-dimensional space but 10% of the two-dimensional embedding space are occupied by neurons).

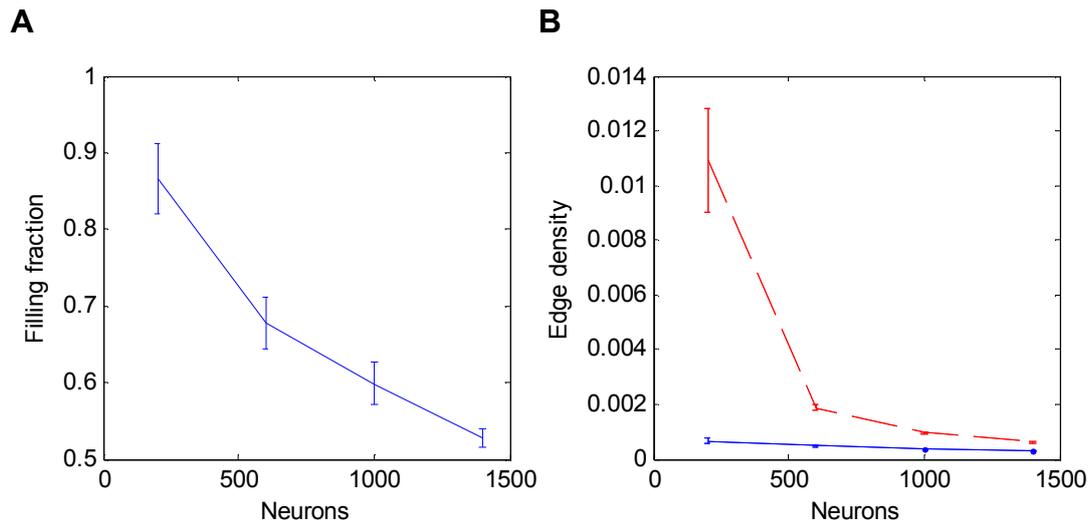

**Figure S3. Dependence of filling fraction and edge density on the number of neurons for three-dimensional growth**. (*A*) Filling fraction, which is the proportion of successfully established connections after axon-cell contact, depends on the number of neurons (vertical bars indicate the standard deviation while lines show the average over 10 generated networks). (*B*) The edge density–the number of existing edges divided by the number of possible edges–decreases for larger numbers of neurons (average and standard deviation as in (*A*); solid line: complete network; dashed line: network without isolated nodes).